\documentclass{PoS}
\usepackage{amsmath,amssymb}
\usepackage{graphicx}

\newcommand*{\MeV}{{\rm Me\!V}}

\def\gtwid{{\,\raise.35ex\hbox{$>$\kern-.75em\lower1ex\hbox{$\sim$}}\,}}
\def\ltwid{{\,\raise.35ex\hbox{$<$\kern-.75em\lower1ex\hbox{$\sim$}}\,}}
\def\leftvec{{\raise1.5ex\hbox{$\leftarrow$}\kern-1.00em}}
\def\rightvec{{\raise1.5ex\hbox{$\rightarrow$}\kern-1.00em}}
\def\half{{\scriptstyle \raise.2ex\hbox{${1\over2}$}}}
\def\threehalves{{\scriptstyle \raise.15ex\hbox{${3\over2}$}}}
\def\third{{\scriptstyle \raise.15ex\hbox{${1\over3}$}}}
\def\third{{\scriptstyle \raise.15ex\hbox{${1\over3}$}}}
\def\twothirds{{\scriptstyle \raise.15ex\hbox{${2\over3}$}}}
\def\fourth{{\scriptstyle \raise.15ex\hbox{${1\over4}$}}}

\def\figref#1{Fig.~\ref{fig:#1}}
\def\Figref#1{Figure~\ref{fig:#1}}

\def\tabref#1{Table~\ref{tab:#1}}

\def\cO{{\cal O}} 

\def\eg{{\it e.g.},\ }
\def\et{{\it et al.}}

\def\inlinetilde{\lower0.8ex\hbox{$\,\widetilde{}\,$}}
\def\chpt{\raise0.4ex\hbox{$\chi$}PT}
\def\schpt{S\raise0.4ex\hbox{$\chi$}PT}
\def\FitQ{{Fit Q}}
\def\FitC{{Fit C}}

\def\burst{{$\times\hspace{-0.30cm}+$}}
\def\circle{{\Large{\raise-0.15ex\hbox{$\circ$}}}}
\def\circle{{\Large{\raise-0.15ex\hbox{$\circ$}}}}
\def\normalbox{{\raise-0.1ex\hbox{$\Box$}}}
\def\prl#1{{\it Phys.\ Rev.\ Lett.\ }{\bf #1}}
\def\prd#1{{\it Phys.\ Rev.\ D} {\bf #1}}

\def\fermilabtwo{{\it Nucl.\ Phys.\ (Proc.\ Suppl.)} {\bf B140} (2005)}

\PoS{PoS(LAT2005)025}

\title{Update on pi and K Physics}

\ShortTitle{Update on pi and K Physics}

\author{\speaker{C.\ Bernard}\\
        Washington University; Saint Louis, Missouri, USA\\
        E-mail: \email{cb@lump.wustl.edu}}

\author{C.\ DeTar, F.\ Maresca, and J.\ Osborn\\
	University of Utah; Salt Lake City, Utah, USA\\
	E-mail: \email{detar@nova.physics.utah.edu},
		\email{maresca@physics.utah.edu},
		\email{osborn@physics.utah.edu}}

\author{Steven Gottlieb and L.\ Levkova\\
	Indiana University; Bloomington, Indiana, USA\\
	E-mail: \email{sg@indiana.edu},
		\email{llevkova@indiana.edu}}

\author{U.M.\ Heller\\
	American Physical Society; Ridge, New York, USA\\
	E-mail: \email{heller@csit.fsu.edu}}

\author{J.E.\ Hetrick\\
 University of the Pacific; Stockton, California USA\\
	E-mail: \email{jhetrick@uop.edu}}
 
\author{D.\ Renner and D.\ Toussaint\\
University of Arizona; Tucson, Arizona, USA\\
	E-mail: \email{dru@physics.arizona.edu},
		\email{doug@klingon.physics.arizona.edu}}

\author{R.\ Sugar\\ 
University of California; Santa Barbara, California, USA\\
	E-mail: \email{sugar@physics.ucsb.edu}}

\abstract{We present an update of the MILC studies of the physics of light pseudoscalars
using improved staggered fermions.  New runs at lighter quark mass, as well as increased
statistics for older sets, are enabling us to improve the results for decay constants
in full QCD.  In addition, we have analyzed quenched runs at two different lattice spacings.
This makes possible a test of the applicability of staggered chiral perturbation theory
in a different context.}

\FullConference{XXIIIrd International Symposium on Lattice Field Theory\\
		 25-30 July 2005\\
		 Trinity College, Dublin, Ireland}

\begin{document}

Since our previously published work \cite{FPI04} on the 
decay constants and masses of the $\pi$-$K$ system, we have continued to generate
and analyze lattices  with $N_f=3$ flavors of dynamical quarks, increasing the
statistics and moving to lighter $u,d$, and (separately) lighter $s$ masses.
We use improved staggered fermions \cite{ASQTAD}.
With the same action, we have also investigated
these quantities in the quenched approximation.
Assuming the chiral fits are good and the continuum extrapolations are accurate,
such calculations make possible:

\vspace{-0.3cm}

\begin{itemize}
\item[{$\bullet$}\ ]{}
A sensitive check of algorithms and methods --- including
the $\root 4 \of {\rm Det}$ trick for dynamical staggered quarks --- 
by comparing full QCD $f_\pi$ to the well-determined experimental value.

\vspace{-0.3cm}

\item[{$\bullet$}\ ]{}
A precise extraction of the CKM matrix element $V_{us}$ from
$f_K$ or $f_K/f_\pi$, competitive with the world-average from
alternative methods.

\vspace{-0.3cm}

\item[{$\bullet$}\ ]{}
A determination of the light quark masses and their
ratios with high lattice precision.

\vspace{-0.3cm}

\item[{$\bullet$}\ ]{}
A test of the applicability of staggered chiral perturbation theory (\schpt) \cite{LEE-SHARPE,SCHPT-CACB}
for describing lattice data --- both $N_f=3$ and quenched.

\vspace{-0.3cm}

\item[{$\bullet$}\ ]{}
An extraction  of the 
 physical coefficients of the $\cO(p^4)$ chiral Lagrangian.

\vspace{-0.3cm}

\item[{$\bullet$}\ ]{}
A determination of the extra, unphysical parameters that enter \schpt. This is important
because these parameters also appear in staggered chiral forms for other
physical quantities, \eg heavy-light decay
constants and form factors \cite{HL-SCHPT}. Fixing the parameters from the
light-light system reduces the systematic errors in heavy-light computations \cite{FERMI-HL}.
\end{itemize}

\vspace{-0.3cm}

In our computations, we analyze two lattice spacings: $a\! \approx\! 0.12\;$fm (``coarse'')
and $a \! \approx\! 0.086\;$fm (``fine'').  
We call the valence quark masses are $m_x$ and $m_y$;
the dynamical quark masses (for $N_f=3$) are
$m'_u\!=\!m'_d\!\equiv\! \hat m'$, and $m'_s$. 
Here primes indicate simulation values; the corresponding masses without primes
are the physical values.
\tabref{lattices} shows the lattice parameters used.

\begin{table}[b]
\begin{center}
\vspace{-0.3cm}
\setlength{\tabcolsep}{3mm}
\begin{tabular}{|c|c|c|c|c|c|}
\hline
$am_q$ / $am_s$  & \hspace{-1.0mm}$10/g^2$ & size & volume & number & $a$ (fm) \\
\hline
0.03  / 0.05   & 6.81 & $20^3\times64$ & $(2.4\;{\rm fm})^3$ & 564 & 0.120\\
0.02  / 0.05   & 6.79 & $20^3\times64$ & $(2.4\;{\rm fm})^3$ & 484 & 0.120\\
0.01  / 0.05   & 6.76 & $20^3\times64$ & $(2.4\;{\rm fm})^3$ & 658 & 0.121\\
0.007  / 0.05   & 6.76 & $20^3\times64$ & $(2.4\;{\rm fm})^3$ & 493 & 0.121\\
0.005  / 0.05   & 6.76 & $24^3\times64$ & $(2.9\;{\rm fm})^3$ & {\bf 400} (197) & 0.120 \\
\hline
{\bf 0.03 / 0.03}  & {\bf 6.79 } &  ${\bf 20}^{\bf 3}\times{\bf 64}$  &  ${\bf (2.4\;fm)}^{\bf 3}$  & {\bf 350 } & {\bf 0.120} \\
{\bf 0.01 / 0.03}   & {\bf 6.75 } & ${\bf 20}^{\bf 3}\times{\bf 64}$   & ${\bf (2.4\;fm)}^{\bf 3}$  & {\bf 349 } & {\bf 0.120} \\
\hline
{\bf quenched } & {\bf 8.00 } &  ${\bf 20}^{\bf 3}\times{\bf 64}$  &  ${\bf (2.4\;fm)}^{\bf 3}$  & {\bf 408} & {\bf 0.119 } \\
\hline
\hline
0.0124  / 0.031   & 7.11 & $28^3\times96$ & $(2.4\;{\rm fm})^3$ & 531 & 0.0863 \\
0.0062  / 0.031   & 7.09 & $28^3\times96$ & $(2.4\;{\rm fm})^3$ & 583  & 0.0861\\
{\bf 0.0031  / 0.031 }   & {\bf 7.08 } &  ${\bf 40}^{\bf 3}\times{\bf 96}$ &  ${\bf (3.4\;fm)}^{\bf 3}$  & {\bf 210 } & {\bf 0.0859 } \\
\hline
{\bf quenched}  & {\bf 8.40 } &  ${\bf 28}^{\bf 3}\times{\bf 96}$ &  ${\bf (2.4\;fm)}^{\bf 3}$  & {\bf 396 } & {\bf 0.0853 } \\
\hline
\end{tabular}

\end{center}

\vspace{-0.5truecm}
\caption{\label{tab:lattices}
Lattice parameters.  Runs and numbers of configuration in normal font
were included in Ref.~\cite{FPI04}; those in bold font are new.  
The lattice sets above the double line are ``coarse;'' those below are ``fine.''
}
\vspace{-0.5truecm}
\end{table}

The relative lattice scale is determined
using the length $r_1$ \cite{R1} from the static quark potential.
We reduce statistical fluctuations in $r_1/a$ by fitting to a
smooth function.
The absolute lattice scale is obtained from the $\Upsilon$ $2S$-$1S$ 
splitting. 
Following the continuum extrapolation in Ref.~\cite{MILC-SPECTRUM},
but using updated HPQCD results \cite{Gray:2005ur}, rather than those
in \cite{PRL}, we obtain
$r_1 = 0.318(7)\;$fm.

As in Ref.~\cite{FPI04} we fit the partially quenched lattice
data to \schpt\ forms \cite{SCHPT-CACB}.
The addition of new runs allows us (sometimes forces us)
to change some details of the fits.
To determine LO 
and NLO chiral parameters 
we fit only
to the low quark-mass region.
The cut on valence quark mass is the same as before:
$am_x+am_y \le 0.021\approx 0.5 am_s$ (coarse) and $am_x+am_y \le 0.017 \approx 0.6 am_s$ (fine).
We now have enough data to cut on sea-quark masses, too: We remove
the $0.03/0.05$, $0.02/0.05$, and $0.03/0.03$ sets for this fit.
Because statistical errors are so
small, (0.1\% to 0.4\% for decay constants, 0.1\% to 0.7\% for squared meson masses), 
we still require NNLO analytic terms in addition to complete
NLO forms to get good fits. 
Such joint fits to decay constants and masses, 
including both coarse and fine lattices, have 26 free parameters: 

\vspace{-0.3cm}

		\begin{itemize}
			
			\item[{$\bullet$}\ ]{} 2 LO parameters: $f$ 
and $\mu$ (decay constant and condensate at tree-level).

\vspace{-0.3cm}

			\item[{$\bullet$}\ ]{} 8 NLO parameters: 4 physical and 2
			taste-violating analytic terms, 2 taste-violating hairpins.

\vspace{-0.3cm}

			\item[{$\bullet$}\ ]{} 10 physical, NNLO analytic parameters.

\vspace{-0.3cm}

			\item[{$\bullet$}\ ]{} 6 tightly constrained parameters 
(prior width = 0.025): give variation of 2 LO and 4 NLO physical parameters with lattice spacing.

		\end{itemize}

\vspace{-0.3cm}

For interpolation around $m_s$, we must include higher quark masses.
Once the LO and NLO parameters are determined, we fix them (up to statistical errors)
and fit to all sea mass sets and wider ranges of valence masses. 
For central values we choose the range $am_x+am_y\le 0.055\approx 1.4m_s$ coarse, and
$am_x+am_y\le 0.0353\approx 1.3m_s$ fine. This fit is called ``\FitC.''
For systematic error tests, the range is widened to  $am_x+am_y\le 0.10\approx 2.5m_s$ coarse, and
$am_x+am_y\le 0.062\approx 2.2m_s$ fine. 
With either of these choices, we need to add in the NNNLO analytic terms (18 parameters, cubic
in quark masses for $f_\pi$ and $M^2_{\pi}/(m_x+m_y)$) to get
good fits.
			
With our old data set, $m'_s$ only changed with $a$, and $m'_s$ was
usually significantly larger than $\hat m'$, so the sea quark mass dependence 
and $a$ dependence were difficult to disentangle.
The new data, which includes
coarse lattices with $am'_s=0.03$ in addition to the previous value $am'_s=0.05$, gives
better control  of the sea quark mass dependence and smaller
$a$ dependence of the LO, NLO, and NNLO parameters.
Including the NNNLO terms and the $a$ dependence of the NNLO terms  gives 28 parameters more than
the low-mass fits described above, for a total of
56 parameters. Twelve of these (LO, LO $a$ dependence, and NLO parameters) are tightly
constrained from the low-mass fits.

\begin{figure}
\begin{center}
\parbox[t]{0.57\textwidth}{
\includegraphics[width=0.57\textwidth]{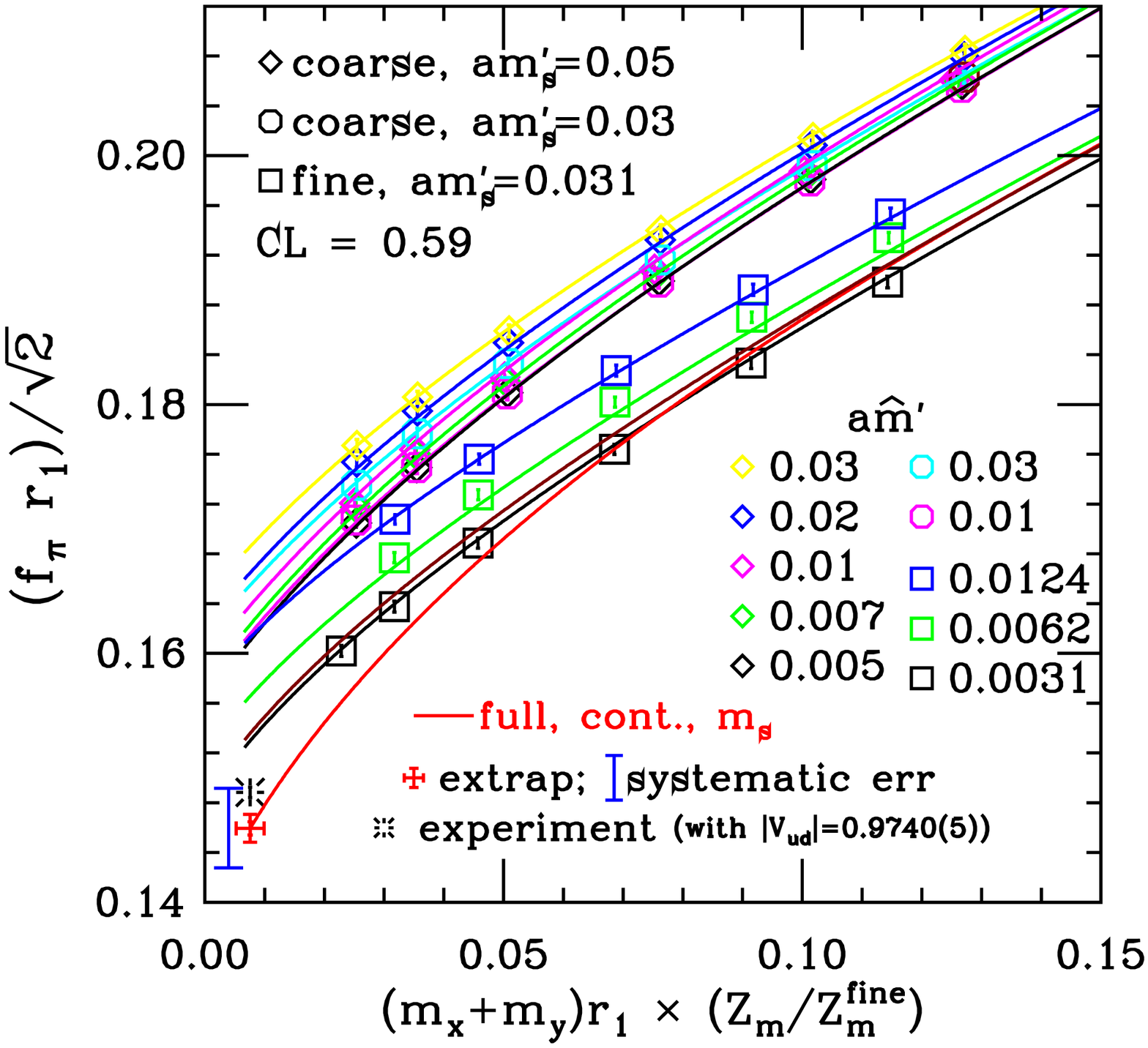}              
}%
\hspace{0.0cm}
\parbox[t]{0.41\textwidth}{
\includegraphics[width=0.41\textwidth]{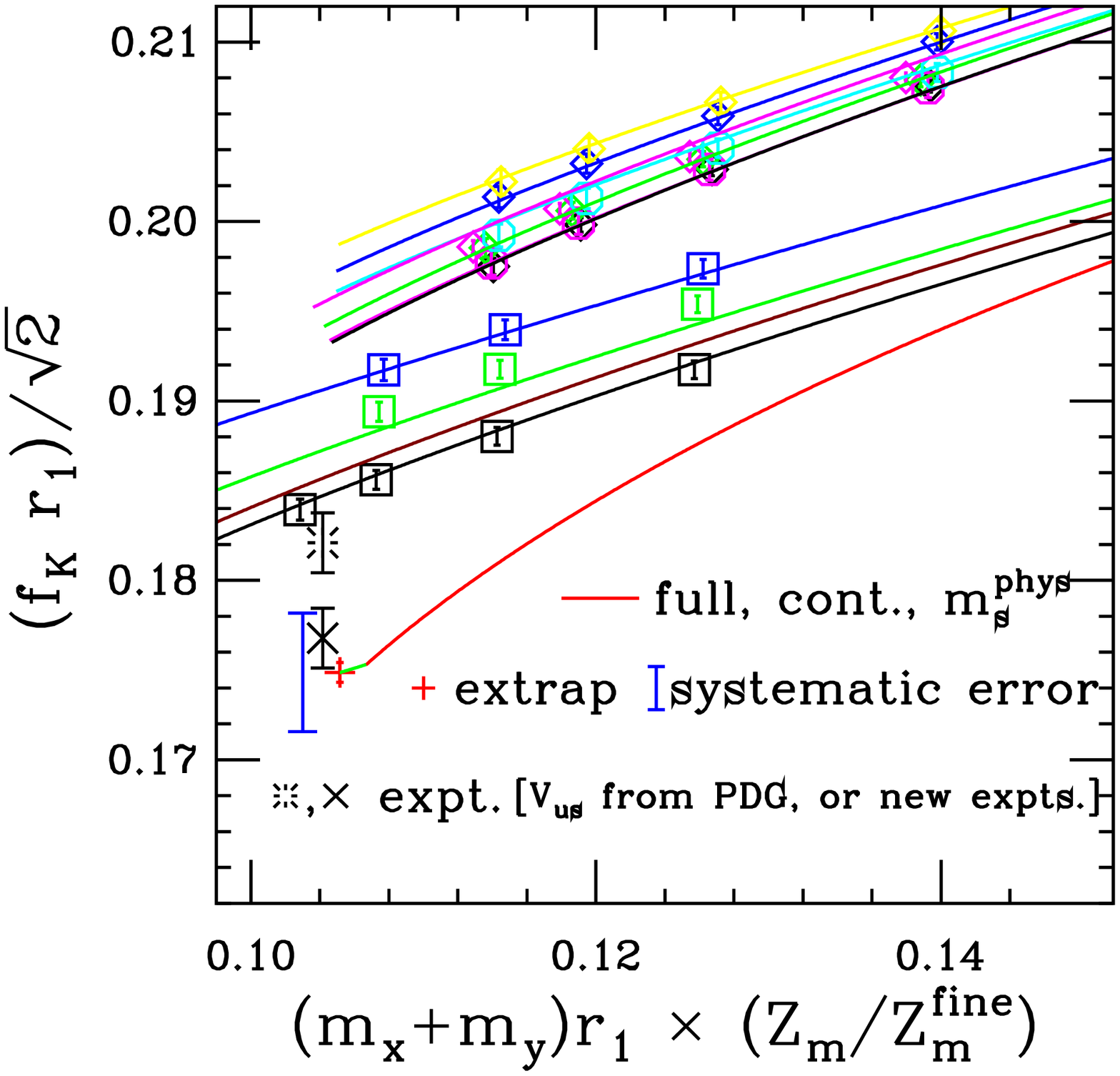} 
}
\end{center}
\vspace{-0.7truecm}
\caption{Left: Comparison of \FitC\ to partially quenched $f_\pi$ data.
Extrapolation to
continuum, setting $m'_s=m_s$,
and going to full QCD gives the red line.  Red $+$
shows the final result after extrapolation $m_x,m_y\!\to\!\hat m$. The maroon line (just
barely visible above the black line)
is the prediction for the 0.0031/0.031 fine run (black squares) based on the other
data.
Right: Same as left, but for $f_K$.
 The short  green
continuation of the  red line keeps light sea masses fixed at  the
average physical value $\hat m$ and extrapolates 
$m_x\!\to\! m_u$.
\label{fig:fpi-fK}}
\vspace{-0.3truecm}
\end{figure}

\Figref{fpi-fK} shows \FitC\ results for $f_\pi$ and $f_K$.
This fit has $\chi^2/{\rm d.o.f.}\!=\!0.99$ for $556$ degrees of freedom (confidence level
CL=0.59). In each plot, the maroon line is the ``prediction'' for
the 0.0031/.031 fine run based on 
a second fit that leaves out that run; it should be compared with the solid
black line that comes from  \FitC.
Since the time of the conference, we have
accumulated about 25\% more 0.0031/.031 configurations, and the effect of
removing or including this run in the fit has decreased. Given that the CL decreases when the run is
removed, we no longer see any reason to consider omitting the run.
We note that the  0.0031/.031 run is still
only about half finished, so there will probably be further noticeable shifts.
In the $f_K$ plot, two ``experimental'' points (shifted slightly
to the left for clarity) are shown. 
Both points are based on the measured leptonic
($K\to\ell\nu$) rate, but \burst\ assumes the PDG value
$V_{us}=0.2200(26)$ \cite{PDG}; while $\times$ assumes the results of
recent experiments $V_{us}=0.2262(23)$ \cite{BLUCHER}.
Both these values of $V_{us}$ come from 
experimental determinations of the semileptonic ($K\to\pi\ell\nu$) rate and
non-lattice theory for form factors.

Our preliminary results for decay constants are:
\vspace{-0.8cm}

\begin{equation*}\label{eq:f_results}
f_\pi  =   128.1 \pm 0.5\pm 2.8\;\MeV\,,\hspace{1truecm}
f_K  =   153.5 \pm 0.5\pm 2.9\;\MeV\,,\hspace{1truecm}
f_K/f_\pi   =   1.198(3)({}^{+16}_{-\phantom{1}5})\ ,
\end{equation*}

\vspace{-0.2cm}

\noindent{where the errors are from statistics and lattice systematics.}
These results are consistent with our previous answers \cite{FPI04}, with slightly
smaller errors.
The current $N_f=3$ results for quark masses are little changed from  those in Refs.~\cite{FPI04,STRANGE-MASS}.

We extract $V_{us}$ from our $f_K/f_\pi$ result. This is probably safer
than using $f_K$ itself, because the ratio is largely free of scale errors.
We obtain $|V_{us}|=0.2242({}^{+11}_{-31})$, which is consistent with 
world-average values, with comparably sized errors. From \figref{fpi-fK}, one can deduce
that using $f_K$ alone would result in a somewhat higher value of $V_{us}$.  The difference comes
from the fact that our $f_\pi$ result is slightly low compared with experiment, although consistent within errors.
Runs planned for the near future, as well as those now in progress,
should allow a further reduction in the errors.

We now turn to the quenched data.
We fit to the same valence mass range as the $N_f=3$ \FitC.
Again, terms through NNNLO are needed; joint fits
to decay constants and masses, including both coarse and fine lattices, have 34 free parameters: 

\vspace{-0.4cm}

	\begin{itemize}
			
			\item[{$\bullet$}\ ]{} 3 LO parameters: $f$, $\mu$, and the quenched chiral
parameter $\delta$ 
\cite{QCHPT}.  We consider $\delta$ to be ``LO'' 
because its effects are not suppressed by powers
of quark mass.

\vspace{-0.3cm}

			\item[{$\bullet$}\ ]{} 7 NLO parameters: 2 physical and 2 taste-violating analytic terms, 2 taste-violating hairpins, and the quenched chiral 
parameter $\alpha$ \cite{QCHPT}.

\vspace{-0.3cm}

			\item[{$\bullet$}\ ]{} 4 physical, NNLO analytic parameters.

\vspace{-0.3cm}

			\item[{$\bullet$}\ ]{} 4 physical, NNNLO analytic parameters.

\vspace{-0.3cm}

			\item[{$\bullet$}\ ]{} 14 tightly constrained parameters 
(prior width = 0.04): give variation of 2 LO, 4 NLO, 4 NNLO, and 4 NNNLO physical parameters with lattice spacing.

\vspace{-0.3cm}

			\item[{$\bullet$}\ ]{} 
2 parameters to allow the $r_1$ scale on the coarse and fine lattices to vary within $1\sigma$. 

	\end{itemize}

\vspace{-0.3cm}

\begin{figure}
\begin{center}
\parbox[t]{0.46\textwidth}{
\includegraphics[width=0.46\textwidth]{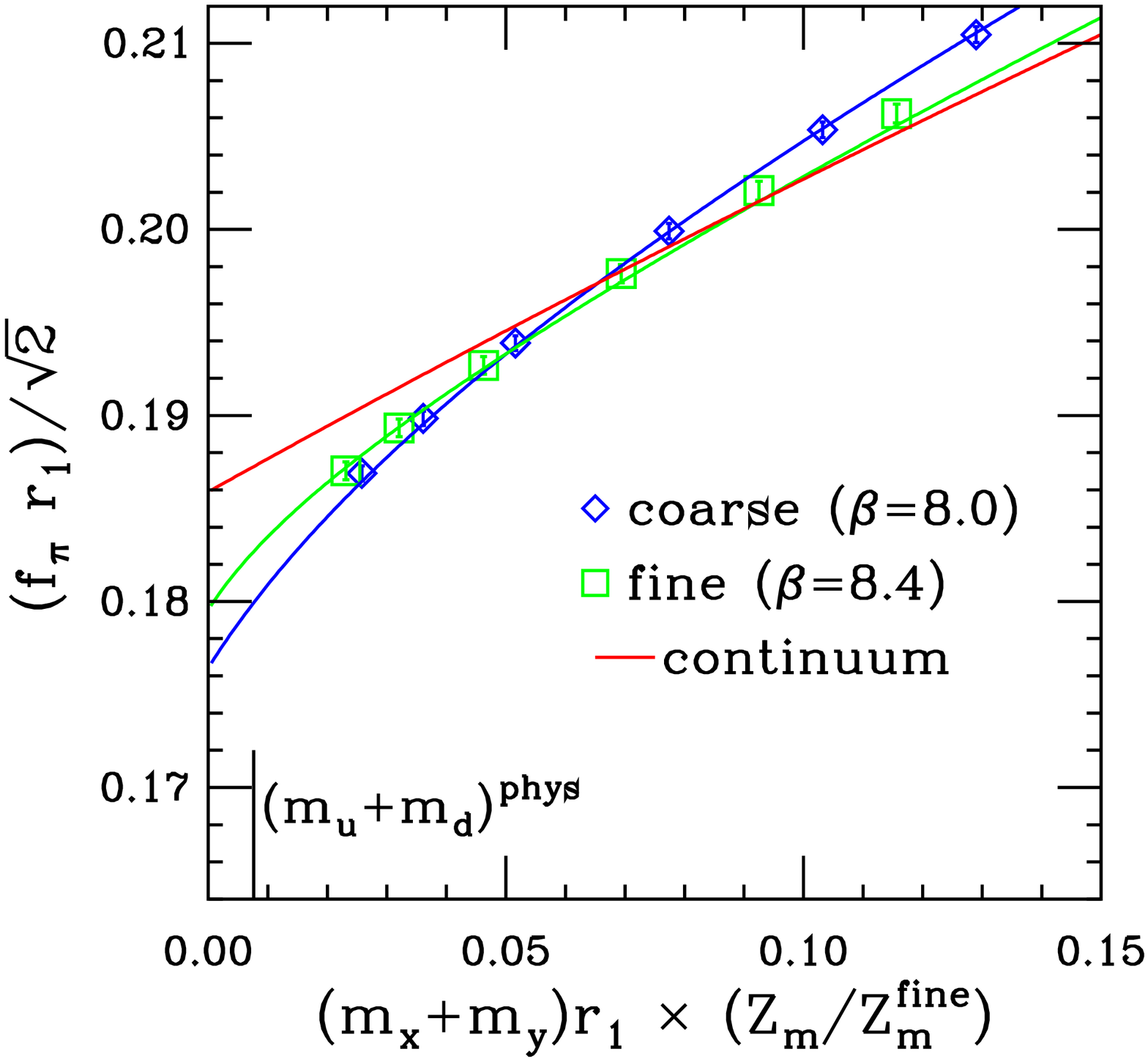}              
}%
\hspace{0.0cm}
\parbox[t]{0.46\textwidth}{
\includegraphics[width=0.46\textwidth]{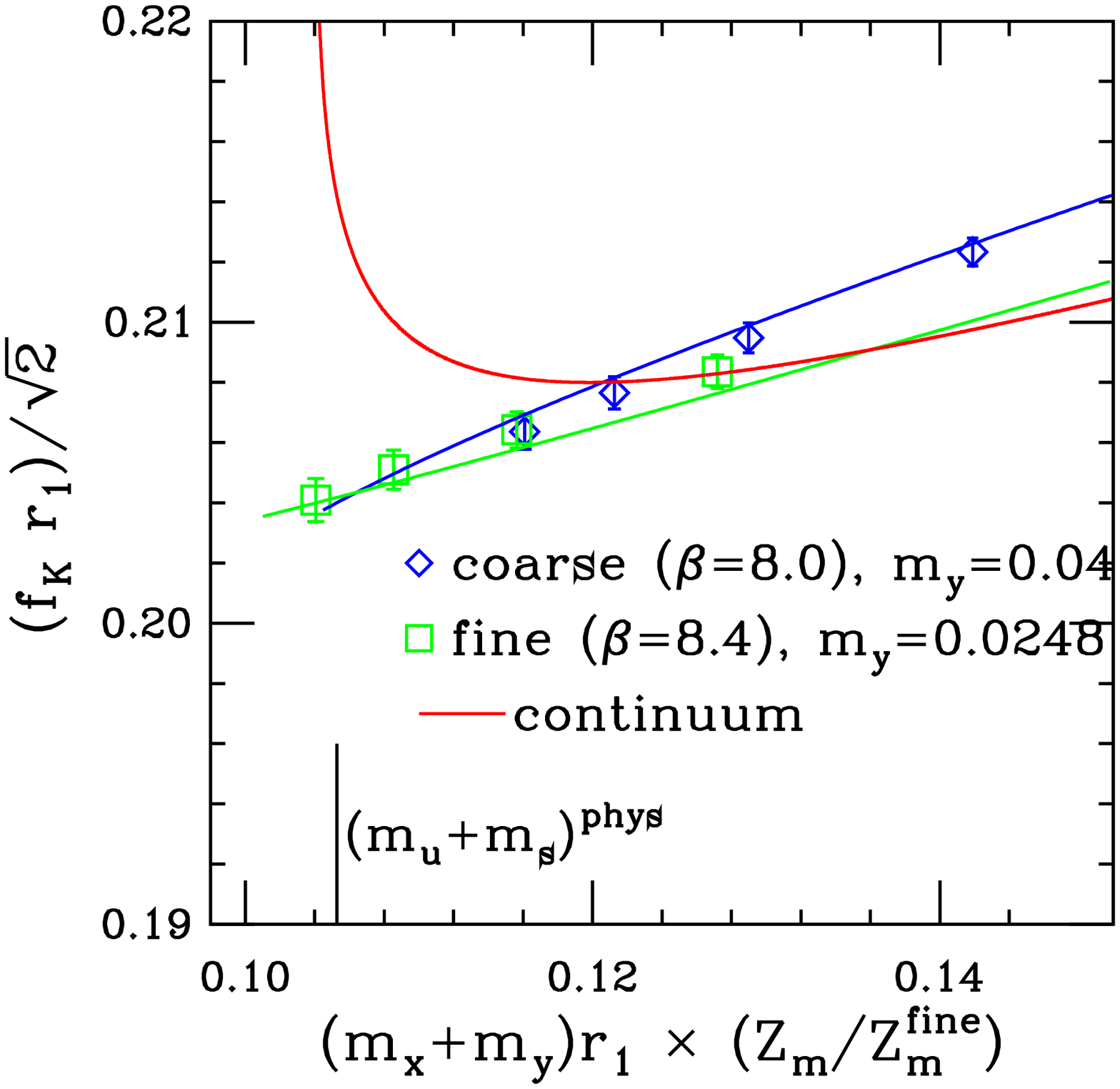}              
}%
\end{center}
\vspace{-0.65truecm}
\caption{Left: Comparison of \FitQ\ to quenched 
$f_\pi$ data. Continuum extrapolation gives the red curve.
$(m_u+m_d)^{\rm phys}$ is from the $N_f=3$ analysis. 
Right:   Same as left, but for quenched $f_K$.\label{fig:Qfpi-fK}}
\vspace{-0.30truecm}
\end{figure}

The parameter $\delta$ multiplies a function of the taste-singlet
mass, which is large ($\gtwid 500\,\MeV$) on coarse lattices because of taste splitting.
The coarse lattices are therefore insensitive
to quenched chiral logs, and $\delta$ is poorly
determined from coarse-lattice fits: $0.02\ltwid \delta \ltwid 0.18$, depending on details of the fit and the higher order
terms included.
Currently, our preferred approach is to obtain $\delta$
from fits to the fine lattices alone, and hold it (as well as the fine lattice
scale) fixed in the joint fits.  
The resulting joint fit, called \FitQ, has 32 parameters and
$\chi^2/{\rm d.o.f.}= 0.96$ for 98 degrees of freedom; CL=0.59.
Comparison of this fit with the quenched $f_\pi$ and $f_K$ data is
shown in \figref{Qfpi-fK}.

The continuum-extrapolated curve (red line)
blows up at low quark mass for $f_K$ 
but not for $f_\pi$: the latter does not
have a 1-loop quenched chiral log in the continuum.
Taking into account \schpt, $f_\pi$
increases significantly with the $a\to0$ extrapolation.  This makes makes our quenched $f_\pi$ differ from 
experiment (or our $N_f=3$ results) by an even larger amount than was reported in Ref.~\cite{PRL}
from coarse data alone.  In fact, we find $f_\pi^{\rm quench}/f_\pi^{N_f=3}\approx 1.28$ (with the
$r_1$ scale). Note, however, that the difference between the raw coarse and fine data for
$f_\pi$ is small, so it is not yet clear how seriously we should take the continuum extrapolation ---
data at smaller $a$ is needed.
This is an even more important issue for quenched $f_K$, where the continuum result blows
up in the chiral limit, but the lattice data is smoothed out by \schpt\ effects.
(This is an example of the lack of commutativity of chiral and continuum limits in \schpt\ for infrared
sensitive quantities \cite{Bernard:2004ab}.)

Values of the taste-violating 
hairpin parameters in the quenched analysis are comparable
to those for $N_f=3$.  We find:

\vspace{-0.8cm}

\begin{eqnarray*}
&r_1^2 a^2 (\delta'_A)^{\rm quench} =  -0.20(2)({}^{+4}_{-8}) \qquad\qquad
&r_1^2 a^2 (\delta'_A)^{N_f=3} =  -0.29(1)(4) \\
&r_1^2 a^2 (\delta'_V)^{\rm quench} =  0.09(5)({}^{+9}_{-7}) \qquad\qquad
&r_1^2 a^2 (\delta'_V)^{N_f=3} = -0.12(2)({}^{+11}_{-5})
\end{eqnarray*}

\vspace{-0.2cm}

{\noindent In both quenched and $N_f=3$, $\delta'_V$ is poorly determined}
and consistent with 0.
For the quenched parameter $\delta$ we are finding: $\delta = 0.09(1)(5)$.
This result is consistent with most quenched evaluations \cite{QUENCHDELTA},
which get $\delta\approx 0.1$,
but not that of the Kentucky group \cite{KENTUCKY}, who obtain $\delta= 0.24(3)(4)$.
(Results in Ref.~\cite{IN-BETWEEN} also tend toward higher values.)
Assuming the analysis in Ref.~\cite{KENTUCKY}, this difference
makes sense.  Despite the low quark masses in the current calculation, the large taste-violations
in the taste-singlet sector puts us in a larger region of meson mass, where the effective
$\delta$ from chiral fits is smaller.  Modulo this issue, which needs further study,
it seems that 
\schpt\ works reasonably well in the quenched theory, as it does in the $N_f=3$ case.


\end{document}